\documentclass[a4paper,final,12pt]{iopart}

\usepackage{iopams}
\usepackage{graphicx}
\usepackage{subfigure}
\usepackage[small]{caption}
\usepackage{epstopdf}
\usepackage{cite}
\usepackage{booktabs}
\usepackage{caption}
\usepackage{tabularx}
\usepackage{float}
\usepackage{url}
\usepackage{xcolor}
\usepackage{multicol}
\usepackage{geometry}
\usepackage{textcomp}
\usepackage{gensymb}
\usepackage{tikz}
\usepackage{comment}

\begin{document}
%


\title[Experimental study of electron power absorption dynamics in RF magnetrons]
{Electron dynamics in planar radio frequency magnetron plasmas: III. Comparison of experimental investigations of power absorption dynamics to simulation results}
\author{B~Berger${^1}$, D~Eremin${^2}$, M~Oberberg${^1}$, D~Engel${^2}$, C~W\"{o}lfel${^3}$,
Q-Z~Zhang${^4}$,
P~Awakowicz${^1}$, J~Lunze${^3}$, R P~Brinkmann${^2}$, and J~Schulze$^{1,4}$}

\address{${^1}$ Chair of Applied Electrodynamics and Plasma Technology, Ruhr University Bochum, Universitaetsstrasse 150, 44801 Bochum, Germany\\
${^2}$ Institute of Theoretical Electrical Engineering, Ruhr University Bochum, Universitaetsstrasse 150, 44801 Bochum, Germany\\
${^3}$ Institute of Automation and Computer Control, Ruhr University Bochum, Universitaetsstrasse 150, 44801 Bochum, Germany\\
${^4}$ Key Laboratory of Materials Modification by Laser, Ion, and Electron Beams (Ministry of Education), School of Physics, Dalian University of Technology, Dalian 116024, People's Republic of China}
\eads{\mailto{berger@aept.rub.de}}
\begin{abstract}

In magnetized capacitively coupled radio-frequency discharges operated at low pressure the influence of the magnetic flux density on discharge properties has been studied recently both by experimental investigations and in simulations. It was found that the Magnetic Asymmetry Effect allows for a control of the DC self-bias and the ion energy distribution by tuning the magnetic field strength. In this study, we focus on experimental investigations of the electron power absorption dynamics in the presence of a magnetron-like magnetic field configuration in a low pressure capacitive RF discharge operated in argon. Phase Resolved Optical Emission Spectroscopy measurements provide insights into the electron dynamics on a nanosecond-timescale. The magnetic flux density and the neutral gas pressure are found to strongly alter these dynamics. For specific conditions energetic electrons are efficiently trapped by the magnetic field in a region close to the powered electrode, serving as the target surface. Depending on the magnetic field strength an electric field reversal is observed that leads to a further acceleration of electrons during the sheath collapse. These findings are supported by 2-dimensional Particle in Cell simulations that yield deeper insights into the discharge dynamics.

\end{abstract}

\vspace{2pc}

\submitto{\PSST}
\maketitle

\begin{section}{Introduction}
	\label{sec_introduction}

The process of depositing thin films is of high importance in a variety of technological applications. A high quality of these films is needed for biomedical and optical applications as well as for the manufacturing of microelectronics \cite{Schmidt2019, Jean2016, Senesky2010, Lieberman2005}. Commonly used in this context is the process of Physical Vapor Deposition (PVD) \cite{Rossnagel2003}. Here, a target is positioned in contact with a plasma at comparably low pressure. This leads to the development of a sheath between the plasma and the target that accelerates positive ions to the target surface. These impinging ions can then sputter atoms from the target that condensate on the walls in the chamber. Putting a substrate in the chamber leads to the deposition of a thin layer of the target material or a compound with the used gas on the substrate surface.\\
This process can be enhanced by adding a magnetic field that is strong close the target surface. This leads to an increase of the charged particle density in this region, which, in turn, leads to an increased ion flux to the target. By doing so, the sputter rate of the target material can be increased, which leads to a shorter process time. A typical magnetic field structure is a torus shaped field that has strong magnetic flux components perpendicular to the target surface at the target edges and at its center while having a strong parallel component in the region in between. Such so-called planar magnetron discharges have the advantage of a higher sputter rate compared to discharges without a magnetic field, but suffer from a decreased degree of target material utilization, since the sputter process predominantly takes place in the magnetized torus, which forms the typical racetrack on the target surface. In order to overcome this disadvantage a cylindrical target or rotating magnets for planar targets can be used \cite{Yeom1989,Yeom1989a,ohtsu_windmillmagnetron_2020}.\\
Depending on the desired process, a wide range of different driving voltage waveforms can be used. A DC power source is used for sputtering of conducting materials, but cannot be used to sputter dielectric material, since the surface charges up when getting in contact with a plasma, which leads to arcing at the target. Even when a metallic target is used, process gas admixtures can form a non-conducting layer on the surface, which results in similar problems \cite{Kelly2000,Braeuer2010,Berg2005,Berg2014,Sproul2005}. This problem can be overcome by pulsing the DC voltage \cite{Kelly2000,Belkind1999,Bradley2009,Britun2014}. Another approach is to use higher frequencies to avoid arcing effects, which allows for the usage of dielectric targets in the deposition process \cite{Chapman1980,Surendra1991,Meyyappan1996}. Commonly used is a radio frequency (RF) of 13.56\,MHz. In such a discharge operated at a low neutral gas pressure of a few pascal or less, the energy of the ions impinging upon the target is determined by the voltage drop across the sheath and typically corresponds to the time averaged sheath voltage, since the mean free path of the ions is larger than the sheath thickness. The voltage drop, in turn, is dominated by the DC self-bias, which can be controlled by changing the applied voltage amplitude.\\
Recently, the Magnetic Asymmetry Effect (MAE) in Capacitively Coupled RF (CCRF) discharges has been investigated both numerically and experimentally \cite{Trieschmann2013,Yang2017,Oberberg2018,Oberberg2019,Oberberg2020}. It was found that process relevant plasma parameters such as the DC self-bias and the ion energy distribution function at the electrodes, which affects the sputter rate at the target as well as characteristics of the deposited films at the substrate \cite{ries_IEDF_2019}, can be controlled by changing the applied magnetic field strength at the powered electrode. This control can be conceptually compared to the Electrical Asymmetry Effect (EAE) in CCRF discharges that allows for a control of plasma parameters by applying a tailored voltage waveform \cite{Heil2008,Donko2008,Schulze2009a,Schuengel2012,Derzsi2013,wang_VWT2DPIC_2021,berger_experimental_2015}.\\
In order to understand the control mechanisms in these discharges a detailed understanding of the electron power absorption dynamics is needed, since these dynamics predominantly affect relevant plasma parameters, namely the energy distribution functions, flux, and density distributions of charged particles and radicals in the discharge. Recent studies presented results of simulations and experiments on how electrons gain energy by the applied power in unmagnetized CCRF discharges \cite{Mussenbrock2008,Schulze2018,wilczek_edynamics_2020}. In the $\alpha$-mode, electrons are accelerated by the expanding sheath \cite{Popov1985,Kaganovich1996,Turner1995,Schulze2008,Wilczek2015}. The $\gamma$-mode is characterized by strong ionization by ion-induced secondary electrons in the plasma sheath \cite{Godyak1986,Schulze2011b,Lafleur2013}. In electronegative discharges the Drift-Ambipolar (DA) mode becomes important \cite{schulze_DA_2011,Brandt2019}, as well as the formation of striations \cite{liu_striations_2017,wang_striations_2019}. Additionally, the electron power absorption dynamics can be affected by the plasma series resonance (PSR) \cite{Czarnetzki2006,Schulze2007a,Wilczek2016} and non-linear electron resonance heating (NERH) \cite{Mussenbrock2006}.\\
The electron heating dynamics in a magnetized capacitively coupled plasma are far less well understood than in the unmagnetized case. Lieberman \textit{et al.} described the enhanced acceleration of electrons in the presence of a magnetic field due to multiple interactions of the particle with the expanding sheath as a consequence of the Lorentz force \cite{Lieberman1991,zhang_magneticbounce_2021,sharma_resonanceHeating_2022}. A study by Turner \textit{et al.} showed that applying a weak magnetic field leads to a heating mode transition from a pressure-heating dominated to an Ohmic-heating dominated discharge \cite{Turner1996}. More recent simulation results by Zheng \textit{et al.} showed that this strong Ohmic heating is decisively enhanced by the Hall current in the azimuthal direction of the discharge \cite{zheng_HallCurrent_2019}. Wang \textit{et al.} recently presented simulation results for a magnetized CCRF discharge that investigate the electron power absorption dynamics as a function of the magnetic field strength in oxygen. It was shown that the power absorption mode changes from DA- to $\alpha$-mode when increasing the magnetic flux density \cite{wang_edynamicsmccp_2020}. Another publication by Wang \textit{et al.} showed that the excitation of the PSR and the NERH at high excitation frequencies is reduced when a magnetic field is applied to a CCP, which leads to a reduction of the averaged plasma density for low magnetic flux densities. When a stronger magnetic field is applied, the plasma density increases again due to a longer interaction time between the electrons and the expanding sheath and an additional acceleration of the electrons in a reversed electric field during the collapsing sheath phase \cite{wang_PSRattenuation_2021}. The aforementioned publications, just as most other publications, assume the magnetic field to be applied solely parallel to the powered electrode's surface.\\
In another recent publication by Zheng \textit{et al.} 2-dimensional-Particle in Cell (PIC) simulations were conducted to investigate the electron dynamics in a magnetron discharge operated in argon. It was found that when using a dielectric target a radially dependent charging is building up at the target surface leading to a reduction of the electric field in the region of the plasma bulk. This, in turn, leads to the formation of a spatially dependent ion energy distribution function at the target \cite{Zheng2021}.\\

In the framework of the companion papers the electron power absorption dynamics in a geometrically asymmetric magnetized CCRF discharge operated in argon are investigated in the presence of a conducting target. While the companion papers \cite{Eremin2021a,Eremin2021b} focus on the fundamental heating mechanisms based on 1d3V and 2d3v PIC/MCC simulations of a single set of typical conditions, in this work the plasma discharge is studied as a function of the neutral gas pressure and the magnetic flux density and based on experiments and 2d3v-PIC simulations. The applied magnetic field has a magnetron-like torus shape that provides closed field lines at the powered electrode. An aluminum disk is attached to the electrode serving as a conducting target. It is found that the magnetic flux density as well as the neutral gas pressure strongly alter the electron dynamics. Firstly, at a higher magnetic flux density and/or higher pressure the electron beam accelerated by the expanding sheath is confined to a region close to the target surface, which leads to stronger ionization in this region, while such a confinement is not observed at low magnetic field and/or low pressure. Secondly, there is a large population of electrons trapped above the racetrack region and remaining in the discharge for a long time. Such electrons are strongly energized due to the generation of the $\vec{E}\times\vec{B}$ drift in the azimuthal direction.
With an increase of the magnetic field this population grows and shifts closer to the powered electrode.


The PIC simulations show that this effect is the strongest in the region where the magnetic field is parallel to the target electrode. A reversal of the electric field is found to be induced by the magnetic field in regions where the magnetic field is perpendicular to the surface.\\
The manuscript is structured in the following way: In section \ref{sec:setup}, the experimental set-up is introduced, followed by a description of the PIC simulations in section \ref{sec:PIC}. In section \ref{sec:results}, the results from both the experiment and the PIC simulations are presented and discussed. Finally, conclusions are drawn in section \ref{sec:conclusion}.

\end{section}

\begin{section}{Experimental set-up}
\label{sec:setup}

\begin{figure}[h!]
	    \centering
	    \includegraphics[width=.7\textwidth]{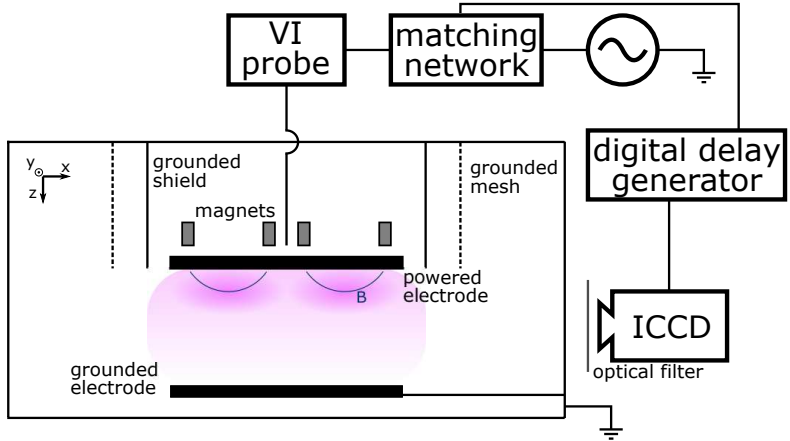}
	    \caption{Schematic of the experimental set-up.}
	    \label{fig:setup}
\end{figure}
	
\noindent	
The experimental set-up used in this work is shown schematically in \fref{fig:setup} and consists of a cylindrical vacuum chamber with a height of 400\,mm and a diameter of 318\,mm. The powered aluminum electrode serves as target, has a diameter of 100\,mm, and is mounted to the top flange of the chamber. A grounded shield and a grounded mesh prevent parasitic plasma ignition towards the grounded chamber walls. A grounded counter electrode is positioned at a gap distance of 70\,mm. The powered electrode system includes NdFeB permanent magnets installed in two concentric rings to create an azimuthally symmetric, balanced, torus-shaped magnetic field. Different magnetic flux densities can be used by using different stacked magnets located behind the target surface. In this way different maximum magnetic flux densities at a reference position 8\,mm below the racetrack, where the magnetic field is parallel to the powered electrode, of 0\,mT (no magnets used), 7\,mT, 11\,mT, 18\,mT, and 20\,mT can be realized. By increasing the distance of the magnets from the electrode surface a magnetic flux density of approximately 5\,mT at the reference position can be facilitated as well. In the direction perpendicular to the target surface the B-field strength decreases exponentially. A detailed description of the configuration as well as measurements of the magnetic flux density can be found in ref. \cite{Oberberg2018}.\\
The discharge is driven by a sinusoidal voltage waveform with a frequency of 13.56\,MHz. A VI probe (Impedans Octiv Suite) is used to measure the driving voltage amplitude $V_0$.\\
All measurements are performed in pure argon (25\,sccm flow rate) as a function of the radial magnetic field strength ($B_0=0-11$\,mT) at the reference position and as a function of the neutral gas pressure ($p=0.5-3$\,Pa).\\
An ICCD (intensified charge-coupled device) camera (Stanford Computer Optics 4 Picos) is used for Phase Resolved Optical Emission Spectroscopy (PROES) measurements. The used objective is equipped with an interference filter with a central wavelength of 750.4\,nm and a full width at half maximum of 1.0\,nm to measure the emission of a transition from the $Ar2p_1$ state. This transition has been used because of the relatively short natural lifetime of the excited state of 22.2\,ns that resolves the period of the RF voltage of 73.7\,ns. Additionally, the threshold energy for the electron impact excitation of this state from the ground state is 13.5\,eV and hence only the excitation by highly energetic electrons is observed. A Digital Delay Generator is used to synchronize the camera with the driving voltage waveform and to allow for a precise control of the camera trigger signal within the applied RF period. The ICCD camera is positioned at a window and measures the emission line-integrated along the line-of sight. Additionally, all images are binned across the radial direction. By doing this all information about radial components of the emission is lost and only the axial dimension is resolved. Considering the radial dependence of the magnetic flux density along the target surface, one should mention that this is quite a severe method of data processing. However, by comparing the experimental results to 2d-PIC simulations all radial effects are investigated in detail by the simulations. Using an Abel transform to calculate radially resolved emission from the line-integrated experimental data was found to lead to a strongly reduced signal-to-noise ratio. The emission is measured with a spatial resolution of about 1\,mm in axial direction and a temporal resolution of 3\,ns. Based on a simplified rate equation model the electron impact excitation rate from the ground state into the observed excited state is calculated. Each plot of the calculated excitation rate is normalized to its respective maximum value. In Ref. \cite{schulze_phase_2010} a detailed description of the diagnostic can be found.

\end{section}

\begin{section}{PIC/MCC code}
\label{sec:PIC}

Here, only a brief description is given. For a more detailed account of the numerical techniques used refer to \cite{Eremin2021b}.

\begin{figure}[h!]
	    \centering
	    \includegraphics[width=\textwidth]{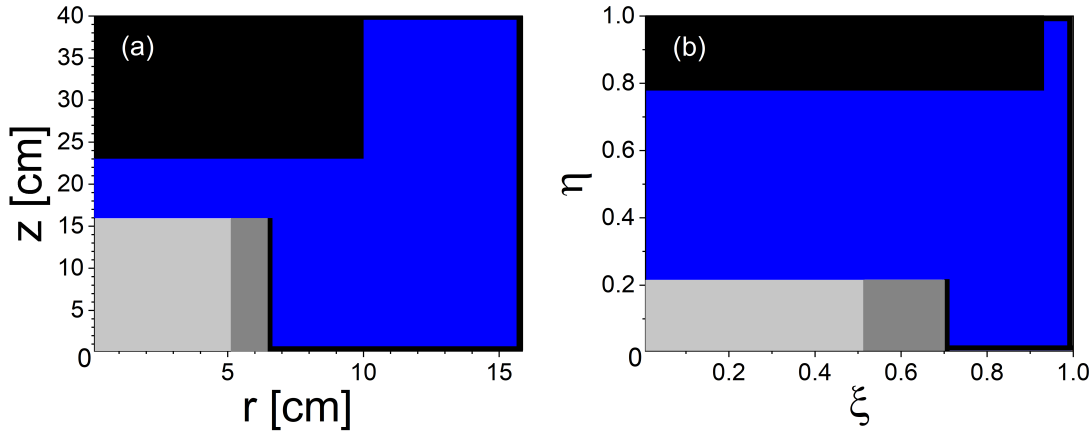}
	    \caption{Modeled geometry in (a) the lab coordinates and (b) the normalized logical coordinates. The blue area denotes the reactor's chamber, while the grounded metal areas are shown in black. The light and dark gray areas mark the powered electrode and the dielectric separator, respectively.}
	    \label{fig:geometry}
\end{figure}

Because of the low pressure used to operate the device, the model should take into account kinetic and nonlocal effects. A suitable algorithm is the particle-in-cell (PIC) method \cite{birdsall_2005,hockney_1988,grigoryev_2002} enhanced with the Monte-Carlo method to model the collisions of plasma particles with neutral particles of the working gas \cite{Birdsall1991}. 

To reduce the computational time, a recently proposed implicit energy-conserving algorithm \cite{Chen2011,Markidis2011} of the PIC method was employed, in a variant discussed in \cite{Eremin2021}. 
Using the energy-conserving PIC algorithm for nonuniform mapped grids \cite{Chacon2013}, one can efficiently reduce the number of computational grid cells even further by allocating fewer grid cells in the areas where no significant physics is expected, and do so without causing any numerical heating. For the model geometry considered in the present work (see Fig. \ref{fig:geometry}(a)), many essential phenomena take place in a small region above the racetrack, whereas the rest of the reactor chamber, by far occupying most of the space, is needed only for a proper account of the DC self-bias. The latter is demanded for a proper description of the plasma series resonance excitation \cite{Oberberg2019,Eremin2021b} and is modeled by including an external network model \cite{verboncoeur_1993,vahedi_1997,schuengel_2012} modified for the implicit energy-conserving PIC method as suggested in \cite{Eremin2021,Eremin2021b}. Furthermore, since for higher magnetic fields (cases with $B_0=7$\,mT and $10$\,mT in section \ref{sec:results}) the powered electrode sheath's width becomes rather small compared to the electrode gap, we have further modified the axial coordinate transformation to ensure an increased resolution in the vicinity of the electrodes. The computational domain in the transformed logical coordinates ($\eta$ in the axial direction z and $\xi$ in the radial direction r) is depicted in Fig. \ref{fig:geometry}(b). For the numerical treatment of cylindrical coordinates, a radially nonuniform computational grid was utilized, where only a few computational cells close to the radial axis were discretized uniformly with respect to $r^2$, whereas the rest of the radial grid was discretized uniformly with respect to $r$. In the logical coordinates we used a uniform grid with $(161 \times 258)$ cells in the radial and the axial directions, respectively. The corresponding Poisson equation in the logical coordinates \cite{Eremin2021a} was solved using the geometrical multigrid algorithm. The nonuniform grid was accompanied by an adaptive particle management algorithm \cite{Lapenta2002,Welch2007,Teunissen2014} designed for the energy-conserving PIC method \cite{Eremin2021b}. Such an algorithm was needed to ensure a balanced resolution of the computational phase space with superparticles and it kept on average approximately $500$ superparticles per cell in the plasma-filled areas in all simulations at the converged state. The time step for the field integration was chosen to be $2.5\times 10^{-11}$\,s and the sub-stepping algorithm was used for the orbit integration to ensure the energy-conservation and the numerical plasma response accuracy \cite{Chen2011}.  
    
Due to the large computational cost the code was parallelized on GPU using a two-dimensional version of the fine-sorting algorithm described in \cite{Mertmann2011}. The method has been benchmarked in 1d \cite{Eremin2021} for a CCP rf discharge in helium \cite{Turner2013} and in 2d for magnetized discharges in the $(\theta,z)$ and $(r,\theta)$ geometries in \cite{Charoy2019} and \cite{Villafana2021}, respectively. The method with all the techniques described above resulted in the the 2d3v energy-conserving implicit electrostatic ECCOPIC2S-M modification of the ECCOPIC code family based on the algorithm of \cite{Eremin2021} and was developed in-house.

The PIC simulations were conducted for argon at $T=450$\,K. Due to the low pressures used, the modeled reactions included elastic scattering, ionization, and excitation \cite{Phelps1999} for the electron-neutral collisions with the elastic scattering and the charge exchange \cite{Phelps1994} for the ion-neutral collisions. The collisions were implemented using the null-collision Monte-Carlo algorithm \cite{Vahedi1995} modified for GPUs \cite{Mertmann2011} using the Marsaglia xorshift128 pseudorandom number generator \cite{Marsaglia2003}, randomly initialized for each thread. The plasma-surface interaction was modeled as follows. The ion-induced secondary electron emission was taken into account with the energy-dependent yield adopted from \cite{Phelps1999} for clean metals under the assumption that the conducting target is sufficiently cleaned by the sputtering process. 
The electron-induced secondary electron emission was modeled in the code after \cite{Sydorenko2006,Horvath2017,Horvath2018}, albeit with non-uniform energy distribution adopted from \cite{Chung1974} and implemented employing the acceptance-rejection method \cite{Nanbu2013}. 
The model parameters were fitted to the measurements of the electron-induced secondary electron emission yield measurements made for Al, which were reported in \cite{Baglin2000THESE} for the mid- and high-energy range and in \cite{Bronstein1969} for the low-energy range. However, it is worth noting that, unlike in dcMS, in rfMS the secondary electron emission is typically not essential for the discharge sustainment \cite{Zheng2021,Eremin2021a,Eremin2021b}. The electron heating mechanisms are much more versatile in rfMS compared to dcMS \cite{Eremin2021b}, but the dominant contribution typically comes from the Hall heating \cite{Zheng2021,Eremin2021a,Eremin2021b}.
\end{section}

\begin{section}{Results}
\label{sec:results}

\begin{figure}[h!]
	    \centering
	    \includegraphics[width=\textwidth]{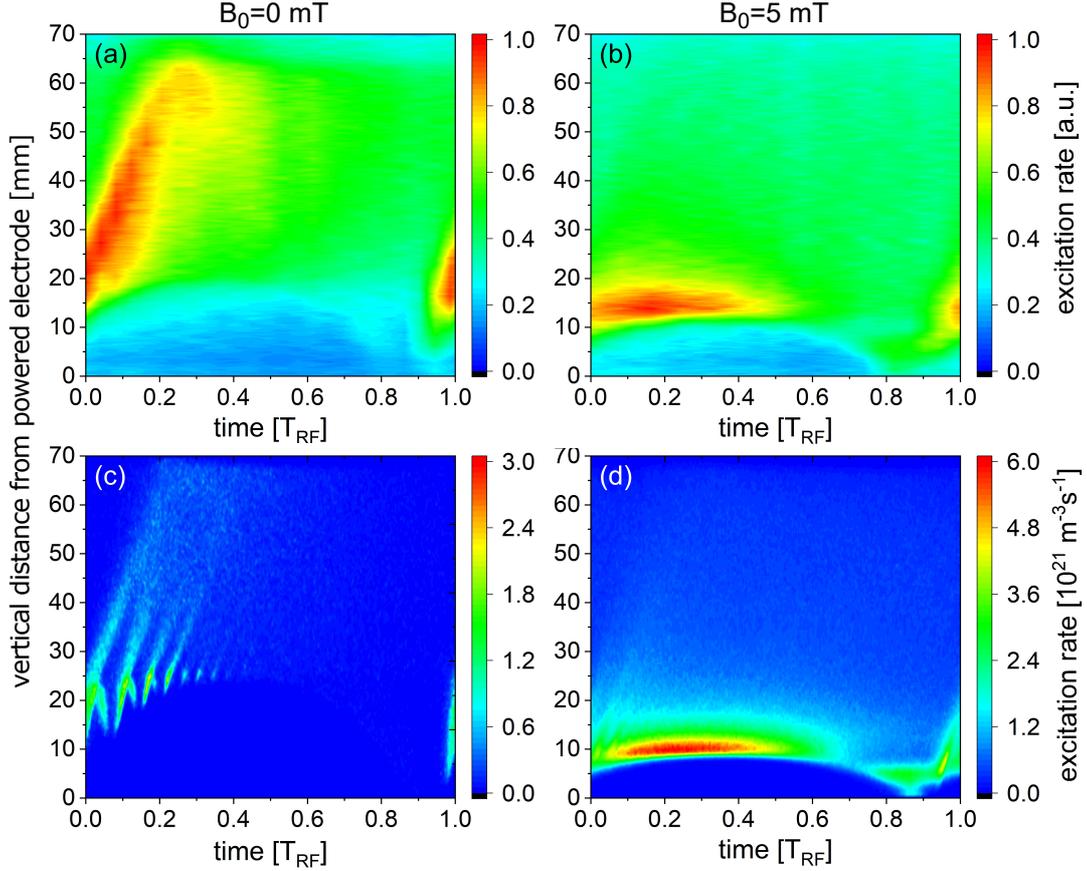}
	    \caption{Spatio-temporal plots of the line-integrated excitation rate from the ground state into the $Ar2p_1$ state with (a) and (c) no magnetic field and (b) and (d) a magnetic flux density of $B_0$=5\,mT at the reference position obtained in (top) the experiment and (bottom) the simulation. Discharge condition: Ar, 1\,Pa, 900\,V driving voltage amplitude.}
	    \label{fig:Exp_1Pa-900V-0,5mT}
\end{figure}

\noindent In Fig. \ref{fig:Exp_1Pa-900V-0,5mT} the results of the PROES measurements at 1\,Pa are shown for two cases: (a) no magnetic field and (b) $B_0=5$\,mT. These plots show the spatio-temporally resolved electron impact excitation rate from the ground state into the $Ar2p_1$ state time resolved within the RF period as a function of the distance from the powered electrode. For comparison the excitation rate obtained from the PIC simulations under the same conditions as in the experiment is presented in Fig. \ref{fig:Exp_1Pa-900V-0,5mT} (c) and (d). When no magnetic field is applied to the discharge the sheath at the powered electrode expands for one half of the RF cycle until it reaches its maximum width of approximately 20\,mm. In the second half of the period the sheath collapses again. According to the experimental results and during the expansion phase an electron beam seems to be accelerated by the sheath that traverses through the discharge and reaches the opposite electrode where it is reflected back into the plasma bulk. This behavior is well known for low pressure capacitively coupled plasmas investigated here \cite{Schulze2008b}. The simulation reveals that actually multiple beams are accelerated during a single sheath expansion phase as previously observed in PIC simulations and in experiments \cite{Wilczek2016,Berger2018}. This behavior cannot be resolved temporally in the present experiment due to the temporal resolution of 3\,ns.

When a magnetic field is applied to the discharge, the excitation dynamics change, as shown in Fig. \ref{fig:Exp_1Pa-900V-0,5mT}(b) and (d). The oscillation of the sheath is still visible but the maximum sheath width is reduced to approximately 10\,mm due to the higher plasma density caused by the magnetic electron confinement. The excitation is still strong during the expansion phase of the sheath but no beam-like structure is visible anymore. The excitation mainly occurs close to the powered electrode in a region where the magnetic field is strong. This can be explained by the reduced electron mobility across the magnetic field lines. One of the electron heating mechanisms is due to the fact that the accelerated electrons gyrate around the magnetic field lines, which guide them back to the sheath region \cite{Lieberman1991}, as investigated before in PIC simulations \cite{wang_edynamicsmccp_2020,zhang_magneticbounce_2021}. In the companion papers \cite{Eremin2021a} and \cite{Eremin2021b} we show that the dominant electron heating mechanism is typically caused by the emergence of a strong and time-dependent $\vec{E}\times\vec{B}$ drift in the azimuthal direction, leading to a force in the azimuthal direction. The resulting Hall heating mechanism \cite{zheng_HallCurrent_2019,Eremin2021a} involves an enhanced time-dependent electric field, which is generated predominantly during the sheath collapse and the sheath expansion and is required to ensure a sufficient level of electron transport across the magnetic field lines \cite{Eremin2021a}. This mechanisms creates a large population of energetic electrons above the racetrack, where they are trapped and move back and forth along the magnetic field lines due to the mirror effect and can remain in the discharge for a long time \cite{Eremin2021b}.

The comparison between the experimental and the computational data shows the good agreement between both results, which allows for a more detailed investigation of fundamental processes by analyzing the simulation data.

\begin{figure}[h!]
	    \centering
	    \includegraphics[width=\textwidth]{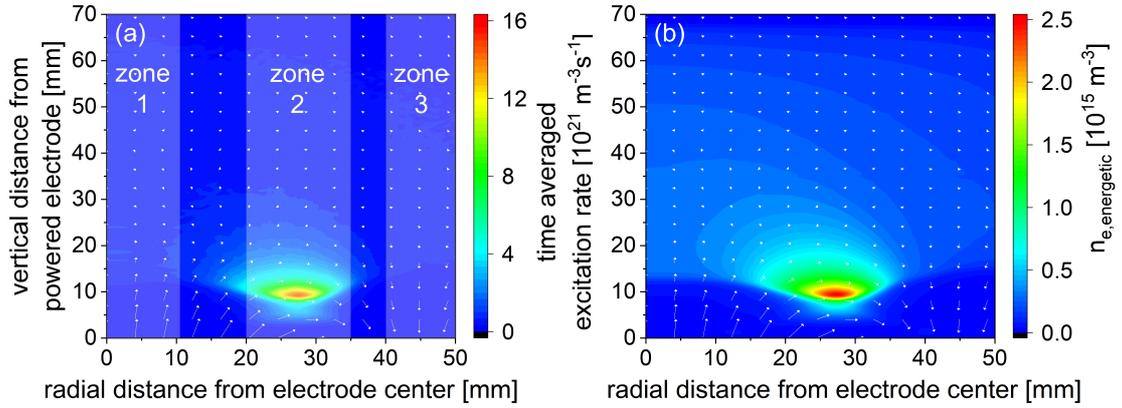}
	    \caption{2-dimensional plots of the (a) time averaged excitation rate from the ground state into the $Ar2p_1$ state and (b) time averaged density of electrons above the ionization energy for argon of 15.8\,eV obtained from the PIC simulation. The boxes in (a) mark three different regions of interest and the arrows mark the magnetic field used in the simulation. Conditions: Ar, 1\,Pa, 900\,V driving voltage amplitude, 5\,mT.}
	    \label{fig:PIC_exc-ne_1Pa-900V-5mT}
\end{figure}

\noindent Fig. \ref{fig:PIC_exc-ne_1Pa-900V-5mT} shows simulation results of the time-averaged excitation rate into the $Ar2p_1$ state and the time-averaged density of highly energetic electrons in subplot (a) and (b), respectively, under the same discharge conditions as before. In contrast to the line-integrated PROES measurements, PIC simulations allow a two-dimensional investigation and, hence, a deeper understanding of the observed effects. The zones indicated in Fig. \ref{fig:PIC_exc-ne_1Pa-900V-5mT}(a) mark three important regions in the discharge. In zones 1 (0-10\,mm of the electrode radius) and 3 (40-50\,mm of the electrode radius) the magnetic field mainly consists of components perpendicular to the powered electrode, while zone 2 (20-35\,mm of the electrode radius) is the region with mainly parallel components. One can see that most of the excitation occurs in zone 2 while in zones 1 and 3 almost no excitation is observed. The same holds true for the density of highly energetic electrons as shown in Fig. \ref{fig:PIC_exc-ne_1Pa-900V-5mT}(b). This means that most of the ionization by electrons in a discharge with such a magnetron-like magnetic field configuration happens in this region as well.

\begin{figure}[h!]
	    \centering
	    \includegraphics[width=0.4\textwidth]{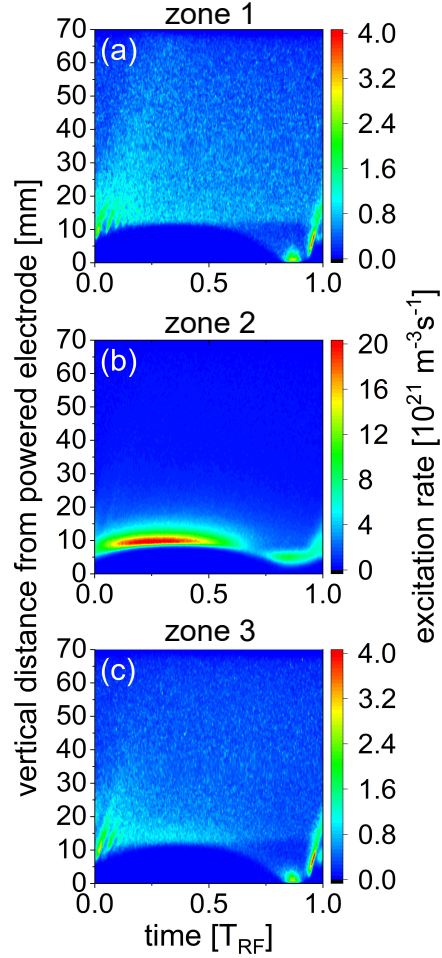}
	    \caption{Spatio-temporal plots of the electron impact excitation rate from the ground state into the $Ar2p_1$ state time-resolved within one RF cycle in the three regions defined in Fig. \ref{fig:PIC_exc-ne_1Pa-900V-5mT}(a) obtained from the simulations. Conditions: Ar, 1\,Pa, 900\,V driving voltage amplitude, 5\,mT.}
	    \label{fig:PIC_exc-xt_1Pa-900V-5mT}
\end{figure}

\noindent In Fig. \ref{fig:PIC_exc-xt_1Pa-900V-5mT} simulation results for the excitation dynamics averaged radially over the assigned regions of interest are shown as a function of time within the RF period and of the axial position between the powered and the grounded electrode. Comparing zone 2 with the results obtained in the experiment one can see a very similar behavior of the accelerated electrons. When the sheath expands starting at t/T$_\mathrm{RF}$=0.9 a strong excitation is visible, which lasts until the maximum sheath width of approximately 8\,mm is reached in the following period at t/T$_\mathrm{RF}$=0.3. This pattern can be associated to electrons being accelerated by the expanding sheath, moving back towards the expanding sheath due to their gyromotion, and hitting the sheath again. As discussed before the limited electron mobility perpendicular to the magnetic field hinders the accelerated electrons from moving across the plasma bulk as it was the case without applied magnetic field. In this way the interaction time of electrons with the expanding sheath is extended due to the presence of the magnetic field. Another strong effect acting on the electrons plays an essential role and is described in detail in the companion papers \cite{Eremin2021a,Eremin2021b}: Due to the large electric and magnetic fields in the region close to the powered electrode, electrons experience a strong ${\bf E}\times{\bf B}$ drift in the azimuthal direction. With an electric field growing in time, this leads to a rapid energization of the electrons, which, in turn, leads to the strong ionization/excitation in zone 2 visible in \ref{fig:PIC_exc-xt_1Pa-900V-5mT}(b).

In zones 1 and 3, however, the excitation pattern changes. In general, the excitation rate in these regions is reduced compared to zone 2. Weak electron beams can be observed when the sheath expands and energetic electrons move into the plasma bulk. The magnetic field here points in the same direction as the electron beam's movement and electrons are not trapped by the magnetic field anymore (see also \cite{Eremin2021b}). Another interesting feature is the increased maximum sheath width of approximately 12\,mm due to the reduced local electron density in these regions. Due to the reduced magnetic field at positions farther away from the electrode, any electron located at the sheath edge is less influenced by the applied magnetic field in zones 1 and 3 compared to zone 2.

\begin{figure}[h!]
	    \centering
	    \includegraphics[width=\textwidth]{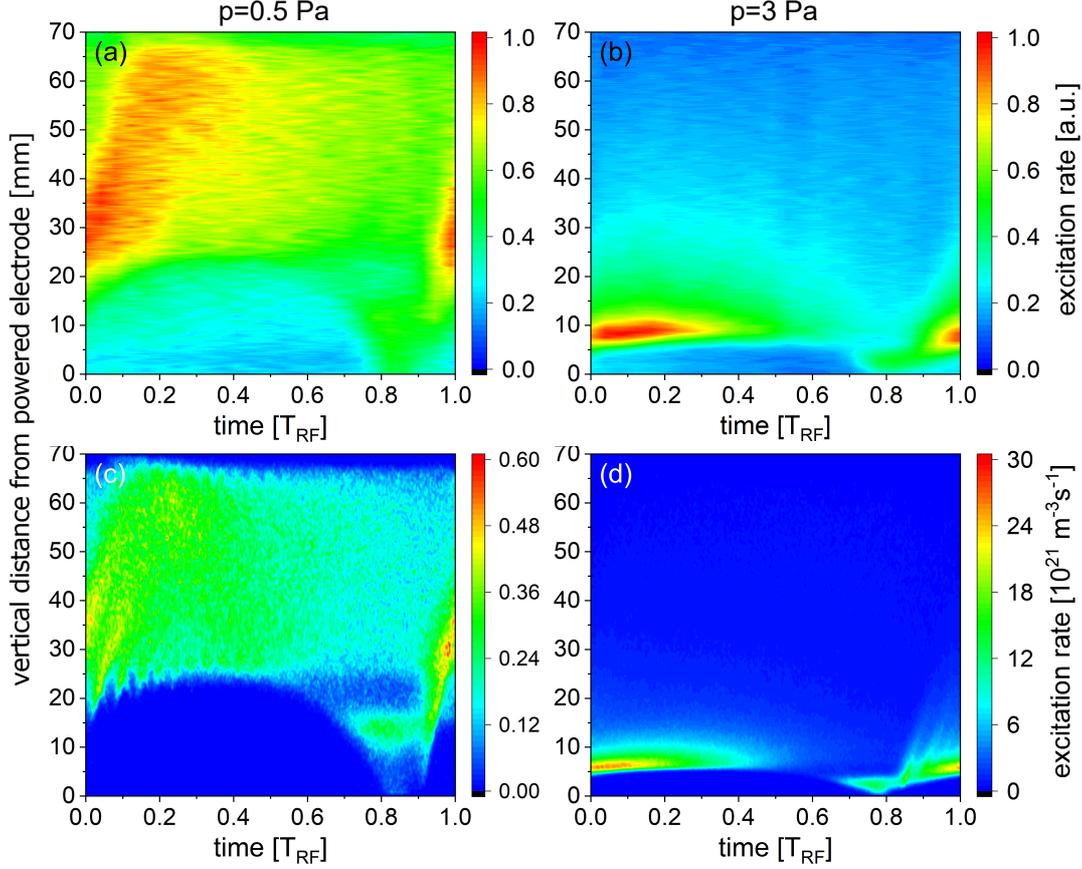}
	    \caption{Spatio-temporal plots of the radially averaged electron impact excitation rate from the ground state into the $Ar2p_1$ state at a neutral gas pressure of 0.5\,Pa [(a) and (c)] and 3\,Pa [(b) and (d)] obtained in (top) the experiment and (bottom) the simulation. Discharge conditions: Ar, 900\,V driving voltage amplitude, 5\,mT}
	    \label{fig:Exp_0.5,3Pa-900V-5mT}
\end{figure}

\noindent In order to further investigate the effect of the magnetic field on the electron dynamics in a CCP, the neutral gas pressure is varied while keeping the magnetic field constant. Fig. \ref{fig:Exp_0.5,3Pa-900V-5mT} shows the excitation rate obtained by PROES measurements and from the simulation. The applied voltage amplitude and the magnetic field are the same as in Fig. \ref{fig:Exp_1Pa-900V-0,5mT}(b). If the pressure is reduced to 0.5\,Pa, the excitation dynamics are strongly altered compared to the case discussed before. Now, the expansion of the sheath leads to an electron beam traversing the bulk and hitting the sheath at the grounded electrode, similar to the dynamics obtained without external magnetic field, as shown in Fig. \ref{fig:Exp_1Pa-900V-0,5mT}(a). When further increasing the neutral gas pressure to 3\,Pa, see Fig. \ref{fig:Exp_0.5,3Pa-900V-5mT}(b), the excitation is localized in a region even closer to the powered electrode compared to 1\,Pa. This behavior is associated to the reduced sheath thickness in case of an increased neutral gas pressure, which is characteristic for CCP discharges. As previously discussed a reduced sheath thickness leads to a higher number of electrons in regions closer to the powered electrode where the magnetic flux density is high and the electrons' Larmor radius is reduced. Hence, the electrons are trapped more efficiently by the magnetic field. On the other hand, when the sheath is thicker at low pressure, electrons are pushed out of the region of strong magnetic field, which allows them to move across the plasma bulk.

\begin{figure}[h!]
	    \centering
	    \includegraphics[width=\textwidth]{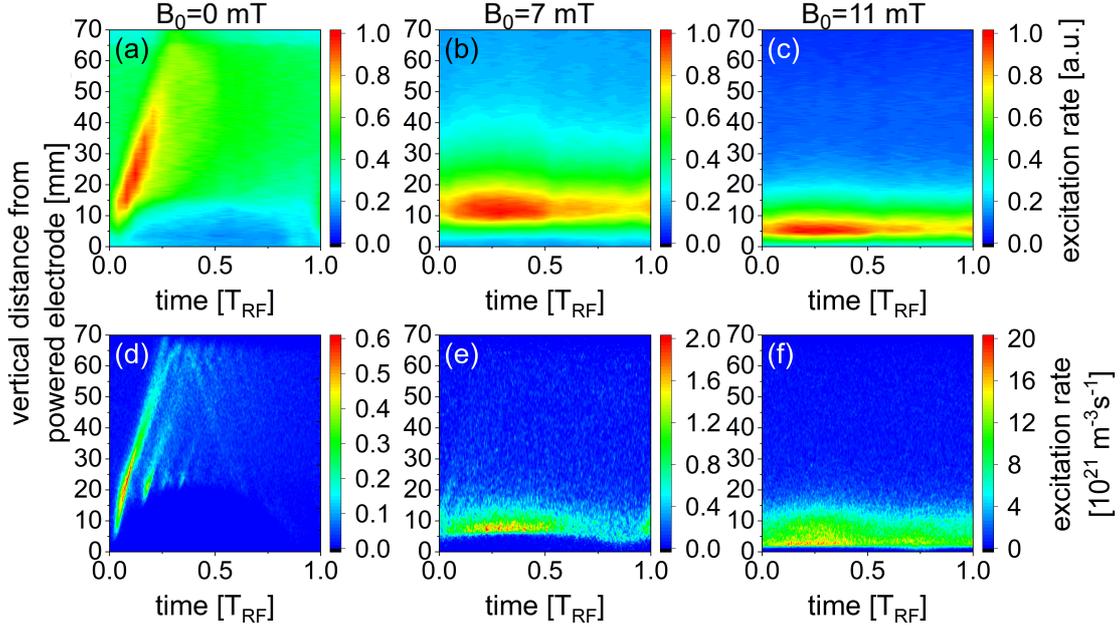}
	    \caption{Spatio-temporal plots of the radially averaged electron impact excitation rate from the ground state into the $Ar2p_1$ state for a magnetic flux density at the reference position of 0\,mT [(a) and (d)], 7\,mT [(b) and (e)], and 11\,mT [(c) and (f)] obtained in the experiment (top) and the simulation (bottom). Discharge conditions: Ar, 1\,Pa, 300\,V driving voltage amplitude.}
	    \label{fig:Exp_1Pa-300V-0,7,11mT}
\end{figure}

\noindent A similar trend can be observed when increasing the magnetic flux density in the discharge. Fig. \ref{fig:Exp_1Pa-300V-0,7,11mT} shows the excitation rate for three different cases at 1\,Pa, i.e. no magnetic field [(a) and (d)], $B_0$=7\,mT [(b) and (e)], and $B_0$=11\,mT [(c) and (f)]. In the experiment a higher power is needed for higher magnetic flux densities in order to ensure a constant applied voltage amplitude. For this reason, the voltage amplitude is reduced to 300\,V in this investigation, both in the experiment and in the simulation, to protect the target from thermal damage. As in the previous case, there are two different mechanism contributing to the observed pattern. Firstly, a highly energetic electron beam forms due to the expanding sheath in the unmagnetized scenario. When increasing the magnetic flux density to 7\,mT, the accelerated electrons are trapped by the magnetic field and strong excitation forms close to the powered electrode. Secondly, the electrons trapped by the mirror effect above the racetrack region and energized due to the Hall heating also reside closer to the powered electrode with the growing magnetic field due a reduced width of the magnetized sheath \cite{Eremin2021b}.


Compared to the electron dynamics at 5\,mT, the sheath width is further decreased. Also, the temporal modulation of the sheath is reduced and the excitation is high for a longer period of the RF cycle. This effect is further enhanced when the magnetic flux density is increased to 11\,mT. This leads to the conclusion that, for most of the RF period, energetic electrons are trapped in a region close to the powered electrode where the magnetic field is strong.

\begin{figure}[h!]
	    \centering
	    \includegraphics[width=\textwidth]{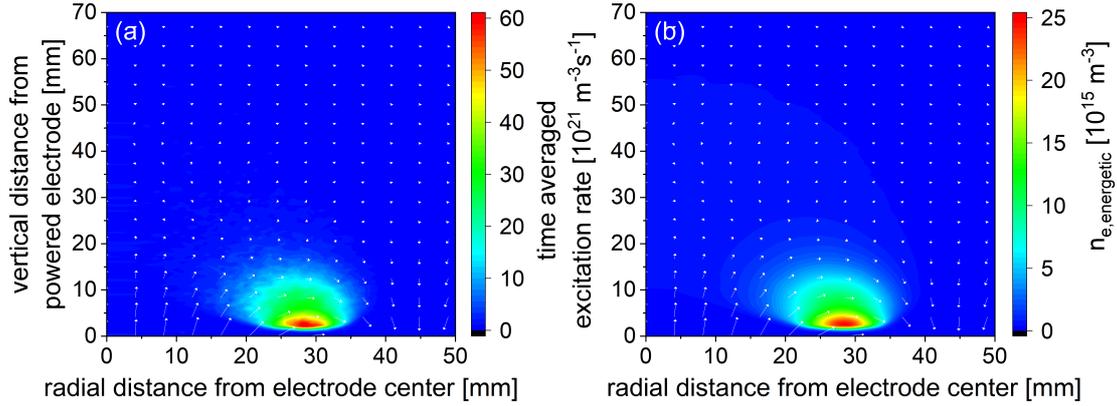}
	    \caption{2-dimensional plots of the (a) time averaged electron impact excitation rate from the ground state into the $Ar2p_1$ state within one RF period and (b) time averaged density of electrons above the ionization energy for argon of 15.8\,eV obtained from the PIC simulation. The arrows mark the magnetic field used in the simulation. Conditions: Ar, 1\,Pa, 300\,V driving voltage amplitude, 11\,mT.}
	    \label{fig:PIC_exc-ne_1Pa-300V-11mT}
\end{figure}

These results are further investigated with the help of the PIC simulations for the case of the strongest magnetic field. The two-dimensional time averaged electron impact excitation from the ground state into the experimentally observed argon state is shown in Fig. \ref{fig:PIC_exc-ne_1Pa-300V-11mT}(a). The pattern is similar to the one shown in Fig. \ref{fig:PIC_exc-ne_1Pa-900V-5mT}. However, the excitation occurs much closer to the electrode surface due to the increased plasma density as a consequence of the enhanced electron confinement in the stronger magnetic field. Even though the voltage amplitude is reduced to one third compared to the case with lower magnetic flux density, the maximum excitation rate is approximately four times higher. The same trend holds true for the time averaged density of energetic electrons shown in Fig. \ref{fig:PIC_exc-ne_1Pa-300V-11mT}(b). Both observations show how efficiently the energetic electrons are trapped by the magnetic field in a magnetized CCP discharge.\\

\begin{figure}[h!]
	    \centering
	    \includegraphics[width=0.4\textwidth]{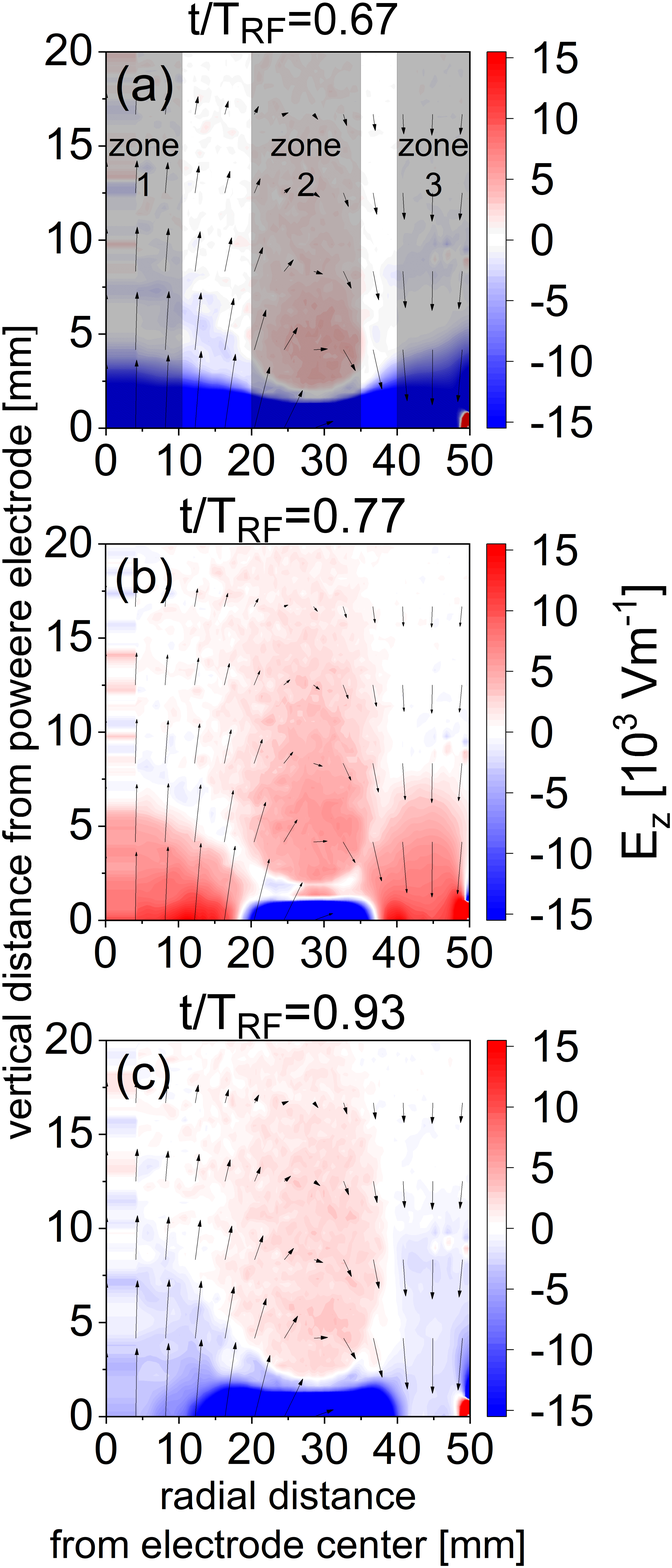}
	    \caption{2-dimensional plots of the electric field component perpendicular to the electrode surface at three different times within one RF period: (a) t/T$_\mathrm{RF}$=0.67 (collapsing sheath), (b) t/T$_\mathrm{RF}$=0.77 (sheath collapse), (c) t/T$_\mathrm{RF}$=0.93 (expanding sheath). The data are obtained from the PIC simulation. The boxes in (a) mark three different regions of interest and the arrows mark the magnetic field used in the simulation. Conditions: Ar, 1\,Pa, 300\,V driving voltage amplitude, 11\,mT.}
	    \label{fig:PIC_Ez_1Pa-300V-11mT}
\end{figure}

Fig. \ref{fig:PIC_Ez_1Pa-300V-11mT} shows the 2d space resolved electric field obtained from the PIC simulations at three different time steps within the RF period again for the strongest magnetic field (11\,mT). The plots only show the region close to the powered electrode up to a distance from the electrode of 20\,mm. It should also be noted that the color scale is chosen to only show values between -15\,kVm$^{-1}$ and +15\,kVm$^{-1}$ to focus on the variations at the sheath edge in contrast to the strong electric field inside the sheath. At t/T$_\mathrm{RF}$=0.67 the sheath at the powered electrode is collapsing, but the electric field is still strong preventing most of the electrons from moving to the electrode surface. In general, the discharge can be divided into three zones again, as marked in Fig. \ref{fig:PIC_Ez_1Pa-300V-11mT}(a). In the center and at the edge of the target electrode, zone 1 and zone 3, respectively, the region of strong negative electric field reaches several millimeters into the discharge, while in the racetrack region, zone 2, the electric field develops in a small region of approximately 1.4\,mm width, which is due to the high local ion density and, hence, the short sheath width in this zone.\\
When the sheath expands again at a later time frame as shown in Fig. \ref{fig:PIC_Ez_1Pa-300V-11mT}(c) a similar behavior can be observed. Investigating the time frame when the sheath is collapsed (t/T$_\mathrm{RF}$=0.77), the electric field strongly alters. In zones 1 and 3, where the magnetic field component perpendicular to the electrode surface is strong, a positive electric field is formed. Such a behavior corresponds to an electric field reversal and leads to the acceleration of electrons towards the electrode surface. This effect is typically observed in electronegative discharges \cite{Vender1992}, in electropositive discharges using tailored voltage waveforms \cite{kruger_voltage_2019}, as well as in previous simulations of CCPs with an externally applied magnetic field parallel to the electrode \cite{Kushner2003,wang_edynamicsmccp_2020} (see also \cite{Eremin2021a}). It usually occurs, if the electron mobility is reduced and the positive ion flux to the surface cannot be compensated by the diffusive flux of thermal electrons anymore. This leads to the generation of an electric field that additionally accelerates electrons to the surface. 

\begin{figure}[h!]
	    \centering
	    \includegraphics[width=0.7\textwidth]{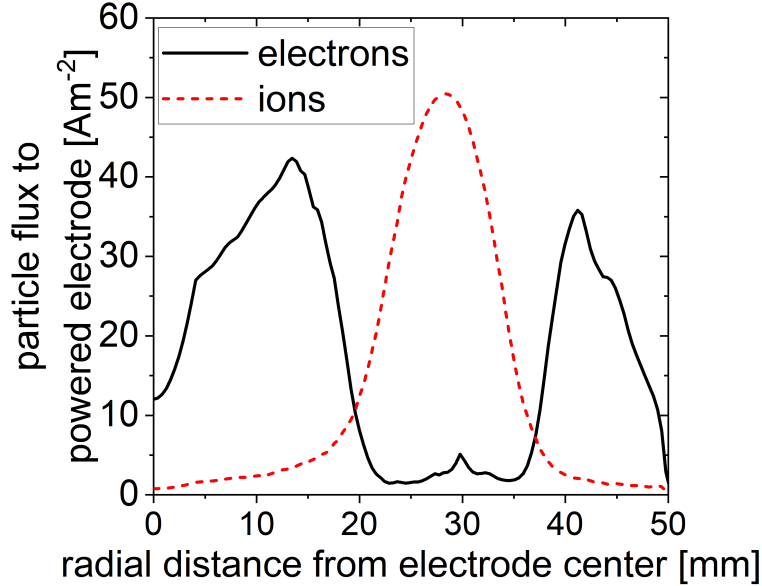}
	    \caption{Time averaged fluxes of charged particles to the powered electrode surface within one RF period as a function of the radial position. The black solid line shows the flux of electrons and the red dashed line shows the flux of positive argon ions. The data are obtained from the PIC simulation. Conditions: Ar, 1\,Pa, 300\,V driving voltage amplitude, 11\,mT.}
	    \label{fig:PIC_fluxes_1Pa-300V-11mT}
\end{figure}

In Fig. \ref{fig:PIC_fluxes_1Pa-300V-11mT} the time averaged fluxes of charged particles to the powered electrode are shown as a function of the radial position on the surface. As expected the ions mainly reach the electrode in zone 2, where the charged particle density is high. In this zone the magnetic field is parallel to the electrode surface. The ions are almost unaffected by the magnetic field and are accelerated by the electric field before reaching the electrode surface. Most of the electrons are not able to reach the surface in this zone due to their reduced mobility across the magnetic field lines, but reach the electrode in zones 1 and 3, where the electron flux is much higher than the ion flux. This means that flux compensation of electrons and ions on time average is happening globally for the target surface with locally very different electron and ion fluxes. This is only possible in the presence of a conducting target, for which a current can flow inside the electrode. In case of a non-conducting target a local flux compensation would be required. In zones 1 and 3, another important mechanism needs to be considered. When an electron enters an area with an increasing density of the magnetic field lines the magnetic mirror force acts on the electron along the magnetic field lines, which leads to a deceleration and ultimately to a reflection of the electron. Here, such a magnetic field configuration can be found in the regions outside of the racetrack above the used magnets, which keeps the electrons from reaching the surface here. In order to compensate the ion flux to the electrode on time average \cite{Schulze2010}, an electric field reversal is, thus, generated in zones 1 and 3 where the magnetic field is perpendicular to the surface, as shown in Fig. \ref{fig:PIC_Ez_1Pa-300V-11mT}(b), to counteract the strong mirror force that acts in the direction away from the target. The effect of the mirror force on magnetized RF discharges is discussed in detail in the companion paper \cite{Eremin2021b}.

\end{section}

\section{Conclusion}
\label{sec:conclusion}
A capacitively coupled argon plasma with an externally applied magnetron-like magnetic field configuration was investigated experimentally by Phase Resolved Optical Emission Spectroscopy measurements to study the electron dynamics in such plasmas. The results are compared to 2d3v-Particle in Cell simulations that offer further insights into the physics of such magnetically enhanced CCPs. It was found that depending on the neutral gas pressure the excitation dynamics of highly energetic electrons differ from those induced by 'classical' sheath expansion acceleration in unmagnetized CCPs. Whereas the free propagation of energetic electron beams generated by sheath expansion heating is found at low neutral gas pressure, comparable to unmagnetized low pressure CCPs, a strong excitation localized close to the target surface is observed when the neutral gas pressure is increased. This is explained by the magnetic confinement of energetic electrons due to a reduced sheath thickness that confines electrons in regions of strong magnetic field close to the powered electrode. The PIC simulations show that this excitation is mainly due to electrons being confined in the region of strong magnetic fields parallel to the target surface, while in the regions of strong perpendicular components electrons are able to leave the magnetized zone. This leads to an acceleration of electrons beams traversing the gap between the electrodes in these regions. A more detailed description of the relevant mechanisms can be found in the companion paper \cite{Eremin2021b}.\\
Additionally, it was found that due to the reduced flux of electrons to the target surface in regions of a strong parallel magnetic field an electric field reversal is generated in regions of a strong perpendicular magnetic field to balance the fluxes of positive and negative particles to the surface globally on time average. This electric field reversal is necessary to overcome the magnetic mirror force acting on the electrons in the regions outside of the racetrack. No local flux balance is observed in the presence of a conducting target, i.e. the ion flux is higher than the electron flux in regions of parallel magnetic field and the electron flux is higher in regions of perpendicular magnetic field.\\
This paper in conjunction with the two companion papers \cite{Eremin2021a} and \cite{Eremin2021b} aids to comprehend the heating mechanisms in RF magnetron discharges. The power absorption dynamics are now understood based on PIC simulations and experimental investigations ranging from fundamental studies in one-dimensional simulations to application-oriented analyses in a real discharge. It was found that the power absorption dynamics look significantly different to unmagnetized CCPs, which is of high importance for industrial applications of these discharges.

\ack
This work was funded by the German Research Foundation within the framework of the Sonderforschungsbereich SFB-TR 87 and the project "Plasmabasierte Prozessf\"uhrung von reaktiven Sputterprozessen" (No. 417888799). \\

\section*{References}
  \bibliographystyle{ieeetr}                                      
  \bibliography{bibliography.bib}     

%
\end{document}